\documentclass[12pt,a4paper]{article}

 \usepackage[dvips]{graphics}
 \usepackage{graphicx}
 \usepackage{afterpage}
 \usepackage{enumerate}
\begin{document}
 %%==========================================================

 \title{\bf Origin of Extremely Asymmetric Stokes V \\ Profiles
in an Inhomogeneous Atmosphere}

 \author{\bf V.A. Sheminova}
 \date{}

 \maketitle
 \thanks{}
\begin{center}
{Main Astronomical Observatory, National Academy of Sciences of
Ukraine
\\ Zabolotnoho 27, 03689 Kyiv, Ukraine\\ E-mail: shem@mao.kiev.ua}
\end{center}

 \begin{abstract}
The formation of unusually shaped Stokes $V$ profiles of the Fe~I
 630.2~nm
line in the solar photosphere are investigated. The
results of numerical 2-D MHD simulation of solar magnetogranulation are used
for this. In their properties, the synthetic unusual  profiles with
extremely asymmetry are similar to the unusual  profiles
observed with a spatial resolution better than $1^{\prime\prime}$
in the network and internetwork regions. According to our results
the unusual profiles mostly appear in clusters along the polarity
inversion lines in the regions of weak magnetic fields with mixed
polarity. As a rule, they are located at the edges of granules and
lanes, and sometimes they are met close to strong magnetic field
concentrations with high velocity and magnetic field strength
gradients. They turned out to appear as clusters in the regions
where large granules disintegrate and new magnetic flux tubes
begin to form. The unusual $V$ profiles may have from one to six
lobes. The one-lobe and multilobe profiles are of the same
origin. The processing causing the extreme  asymmetry of the 
profiles are characterized by one or several polarity reversal
along the line of sight as well as by complicated  velocity and
field strength gradients. The greater the number of profile lobes,
the greater is the probability of the field gradient sign change.
Hence it follows that the magnetic field should be very
complicated in the regions of formation of extremely asymmetric
$V$ profiles. This is  confirmed by immediate results of MHD
granulation simulations, which demonstrate the formation of
vortices and turbulence by the velocity shear at down draft edges.
These processes add complexity to the magnetic field structure by
mixing field polarities, particularly at the edges of
granules.
\end{abstract}
%-------------------------------------------------

\section{Introduction}
     \label{S-Introduction}

The profiles of circularly polarized radiation or Stokes $V$
profiles of absorption lines are used  most often to study the
magnetic field structure and the motion of matter in magnetic
features on the Sun. They are observed not only at the locations
of strong magnetic concentrations in the plages and networks but
also in the quiet Sun, where weak internetwork fields prevail.
According to polarimetric observation data \cite{14,25,27,28,30},
the Stokes $V$ profiles (hereinafter referred to as ''profiles``)
in active and quiet regions on the Sun are variable in shape. 
In dynamic inhomogeneous atmospheres the profiles are almost without
exception asymmetric due to the gradient of the magnetic and
velocity fields along the line of sight \cite{31}. Depending on
their complexity, the profiles are said to be usual (two lobes of
opposite signs, Fig. 1) or unusual (three and more lobes or a
single lobe, Fig. 2).
%%%%%%%%%%%%%%%%%%%%%%%%%%%%%%%%%%%%%%%%%%% Figure 1
  \begin{figure}
 \centerline{\includegraphics [scale=0.8]{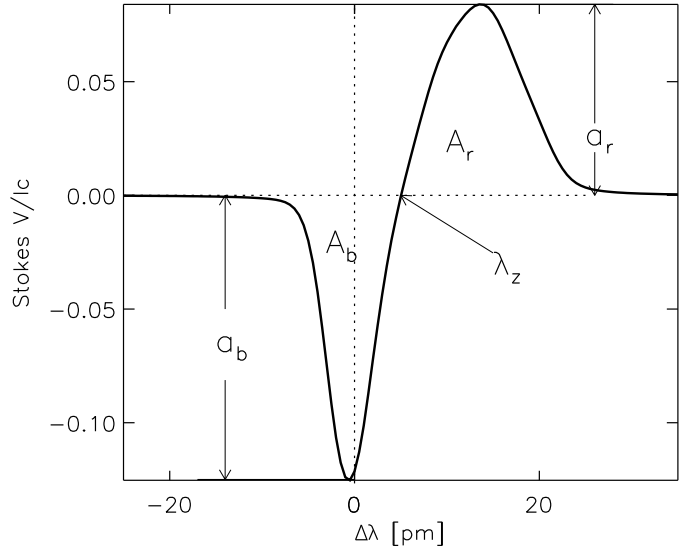}}
 \hfill
 \caption
{Usual asymmetric $V$ profile of the Fe I 630.2~nm  line 
 synthesized inside a strong flux tube of negative
polarity. The profile shape is described by the amplitudes
$a_b,~a_r$, and areas  $A_b,~A_r$ of the blue and red wings, by
the zero-crossing  $\lambda _z$  and by the amplitude and area
asymmetry parameters $ \delta a = \frac{|a_b|-|a_r|}{|a_b|+|a_r|}$
and $ \delta A = \frac{|A_b|-|A_r|}{|A_b|+|A_r|}$. } \label{V-1}
 %}
 \end{figure}

%%%%%%%%%%%%%%%%%%%%%%%%%%%%%%%%%%%%%%%%%%% Figure 2
 \begin{figure}
 \centerline{\includegraphics [scale=1.]{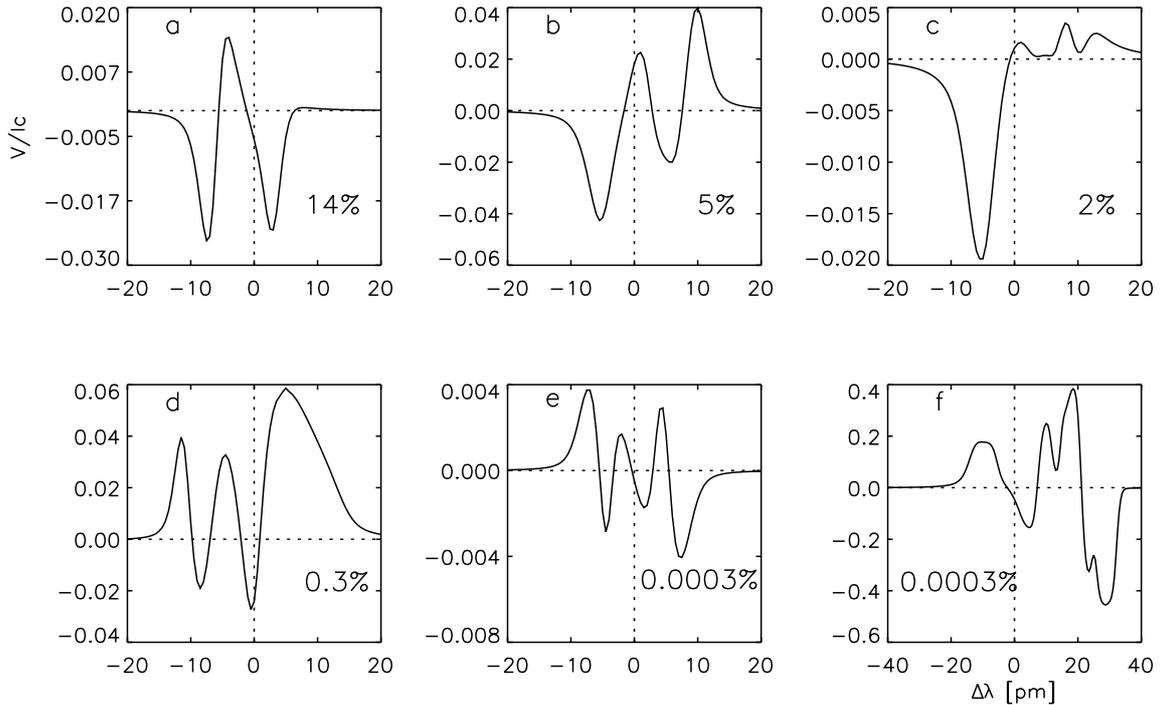}} \hfill
 \caption
{Typical shapes  of unusual 
$V$  profiles of the Fe I  630.2~nm  line. } \label{V-8}
 %}
 \end{figure}

The asymmetry of usual profiles has been studied quite well
 \cite{10,14,18,28}. Outside sunspots at the disk center the majority of
usual profiles have positive asymmetry of amplitudes and areas,
i.e., the blue wing is stronger than the red one; the amplitude
asymmetry is greater, on the average, than the asymmetry of areas.

Unusual profiles were first found in sunspot penumbrae  \cite{16}.
Their origin was attributed to unresolved fine magnetic structures
with opposite polarities and to various flows of matter in them
 \cite{32}. Later on, unusual profiles were observed in active regions
 \cite{21,24}, they were most often met along the neutral line. They
were explained by siphon flows  \cite{22} or by a superposition of
two or more magnetic components of different polarities in the
same spatially resolved magnetic element  \cite{8}. In addition to
the two-component or three-component models with magnetic fields
of opposite polarities, the idea of magnetic field
microstructuring \cite{24,25} was also used. Within the scope of
this idea the magnetic field in plages was regarded as a
statistical of very small, optically thin, nearly vertical
magnetic elements frozen in a nonmagnetic medium with downflows.
Using the inversion method, the authors of  \cite{24,25}
successfully reproduced various unusual profiles observed in
network and internetwork fields. Some important results were
obtained by simulating various situations in the behavior of
atmospheric parameters. Some specific magnetic and velocity field
configurations which are separated in space and which produce the
observed extremely asymmetric profiles were discussed in
\cite{11}. Joint influence of the magnetopause and the temperature
inversion on profile shape can  be a cause of strong asymmetry
\cite{34}.

A considerable number of unusually shaped profiles (about 3.5
percent) in quiet regions were first found in observations with a
spatial resolution better than $1^{\prime\prime}$ \cite{28}.
Analysis of these observations  \cite{27} revealed that those
unusual profiles mostly appeared in clusters in bipolar active
regions near the neutral line, where strong linear polarization is
observed. To account for the complicated line shapes, the authors
of  \cite{27,28} supposed that they are formed in the
superposition of usual profiles with opposite polarities (the
so-called mixed-polarity profiles). The appearance of
mixed-polarity profiles far from the neutral line was explained by
the presence of small magnetic bipolar loops  \cite{17,33} or
highly inclined fields  \cite{32} in the photosphere. Such
profiles can also appear due to the turbulent  \cite{7} or
granulation  \cite{13} character of the magnetic field generated
by local granulation flows; in this case multiple polarity
reversal are also possible in magnetic internetwork fields.
Observations  \cite{14,30} confirmed the presence of
unusual profiles in quiet regions and found  that their fraction  depends
on spatial resolution, spectral line, and magnetic flux. A decrease 
in the magnetic flux results in a greater fraction of unusual profiles.

Thus, several mechanisms were proposed for the extreme asymmetry
of the Stokes $V$ profiles. They can be tested through a
sophisticated treatment of the Stokes profiles with a very high
spatial resolution such that the photospheric surface structures
may not affect the profiles.  The highest resolution 
profiles can be obtained only in
the numerical MHD simulation of magneto-convection  \cite{26}.
Analysis of such profiles allows to study the causes of the unusual profile
asymmetry.  It needs no
scenarios of unusual profile formation, and it proposes a model
atmosphere which fits the solutions of MHD equations. This
approach is successfully used in the solar physics
 \cite{12,15,20,23,6,35}, it proved to be very fruitful in the
investigation of the Stokes profile formation, and a wide variety
of synthetic unusual profiles were obtained.

In this paper we continue our research  \cite{20}  on the origin of unusual profiles
using the synthetic Stokes profiles and the results of the 
2-D MHD models \cite{9}.

%-------------------------------------------------

\section{The MHD models}
     \label{MHD models }

The inhomogeneous atmosphere parameters necessary for calculating
the profiles were derived by Gadun  \cite{3} in the 2-D MHD
simulation of magneto-convection. The simulated region dimensions
were 3920$\times$1820 km with a spatial step of 35 km. The atmospheric
layer extended over about 700 km. The initial model was
nonmagnetic, two-dimensional, and hydrodynamical. The initial
magnetic field was taken in the form of a loop with strength
varying with depth. The field strength averaged over the entire
calculation region $ <|\rm {B} |>$ was equal to 5.4~mT, with the
upper and lower boundary conditions $B_x = 0$ and $\frac{\delta
B_z}{ \delta z} = 0$. The peculiarity of the adopted initial field
was that the field and the initial MHD model atmosphere parameters
were not self-consistent at the initial moment, so that the
initial field was chosen such that it might correspond in the best
manner to the topology of flows and the equipartition conditions
in the simulated region.

At the beginning of simulation the initial field and the diffusion
field (which spreads into the simulation region) accumulate at the
simulation region base, where the field strength should be greater
by the equipartition condition. The possible mechanism for the
magnetic field accumulation are the topological pumping caused by
density gradient  \cite{2} and the local dynamo in the upper
turbulized layers of the convective zone  \cite{7}. The dynamo
mechanism is activated when the kinetic energy exceeds the
magnetic one and the supply of the magnetic energy can be
replenished. Figure 2 in  \cite{3} shows that the kinetic energy
prevails for a long time in the subphotospheric layers in the MHD
models we use, and the magnetic field seems to be generated there
through the dynamo mechanism. At the same time, in the atmospheric
layers this mechanism works for the first 20 minutes only and the
magnetic field distribution is mainly chaotic (turbulent) at that
time.

Concurrently with the dynamo mechanism, the kinematic mechanism
also works in the simulation region -- the magnetic field is
pushed out from the central parts of convection cells by ascending
convective flows, and the field lines are concentrated at the cell
edges, where downflows are located. This mechanism remains
efficient until the field achieves the equipartition level. The
kinematic mechanism in the solar photosphere can provide
concentrations of 70--100 mT fields at the level of radiating
layer in the region of downflows.

Next the thermal mechanism, or convective collapse, comes into
effect -- the magnetic field concentrated by the kinetic mechanism
prevents reverse flows from penetrating into the magnetic
concentration, and thus the concentration becomes supercooled and
descends into deeper layers, causing the magnetic field force line
tension. A low-pressure region is formed in the upper part of this
magnetic configuration, oscillations arise in the configuration as
a result, and it becomes unstable. The principal peculiarity of
this mechanism is that it depends on the size of downflow region
-- the wider the region, the higher is the probability for
convective collapse to develop.

There is one more factor which facilitates the concentration of
vertical magnetic field in the photospheric layers -- the surface
mechanism which begins to operate in the course of fragmentation
of ascending large-scale convective cells  \cite{4,9}. Horizontal
surface fields are captured by descending plasma at the granule
center and are carried to deeper layers, where they form compact
magnetic plumes intensified by the thermal effect.

By the action of all the above mechanisms over a simulation period
of 50 min, the magnetic field attains the strength $<|\rm {B}|>
\approx  40$ mT. The strong magnetic concentration (flux tubes)
thus formed begin to act on the surrounding plasma -- bright dots
are observed, oscillating ascending/descending flows appear in
intergranular lanes. The granulation pattern also changes. The
magnetic field has a stabilizing effect on granulation -- the
granules themselves and their horizontal shear displacements
become smaller. Kilogauss magnetic flux tubes are formed and
develop further in the simulation region. Sometimes flux tubes of
opposite polarities annihilate and tubes of the same polarity
merge. Areas between flux tubes are filled with nearly horizontal
weak magnetic fields of mixed polarity. The mean field strength
varies from 40 to 50 mT over the whole simulation region.
%-------------------------------------------------

\section{Synthetic $V$ profiles and their classification}
     \label{Profile }

For this study we took a one-hour long sequence of MHD models
after 60 min of simulation. The time step is 60 s before the
moment 94 min and 30 s after it. We used 86 snapshots of
the simulated atmosphere. Each snapshot has 112 columns, the
Stokes profiles of the Fe~I line $\lambda$~630.2 nm were
calculated for each column by integrating the Unno-Rachkovskii LTE
equations. In all, 9632 profiles were obtained. These profiles
correspond to those observed in the quiet network and internetwork
at the solar disk center, where there are strong fields of
magnetic flux tubes of about 150~mT and weak granule fields from
70 to 1~mT at the level $\tau_ 5=1$. 

The neutral iron line $\lambda$~630.2~nm is best suite for our
calculations. Its Stokes profiles are highly sensitive to the
magnetic field and velocity field gradients in the solar
photosphere at levels from 0.0 to $-$4.0 on the logarithmic scale
of optical depths. The mean depth of the $\lambda$ 630.2 nm line
formation along columns in inhomogeneous models can vary from
$-$0.5 to $-$1.5 depending on model parameter variations, and
therefore it is hard to specify some average level of the
formation of line profiles in an inhomogeneous atmosphere. It
should also be noted that solar polarimetry observations are most
often made in the $\lambda$~630.2~nm line
\cite{10,15,23,24,25,28,30}.

 %%%%%%%%%%%%%%%%%%%%%%%%%%%%%%%%%%%%%%%%%%% Figure 3
 \begin{figure}
 \centerline{\includegraphics    [scale=0.95]{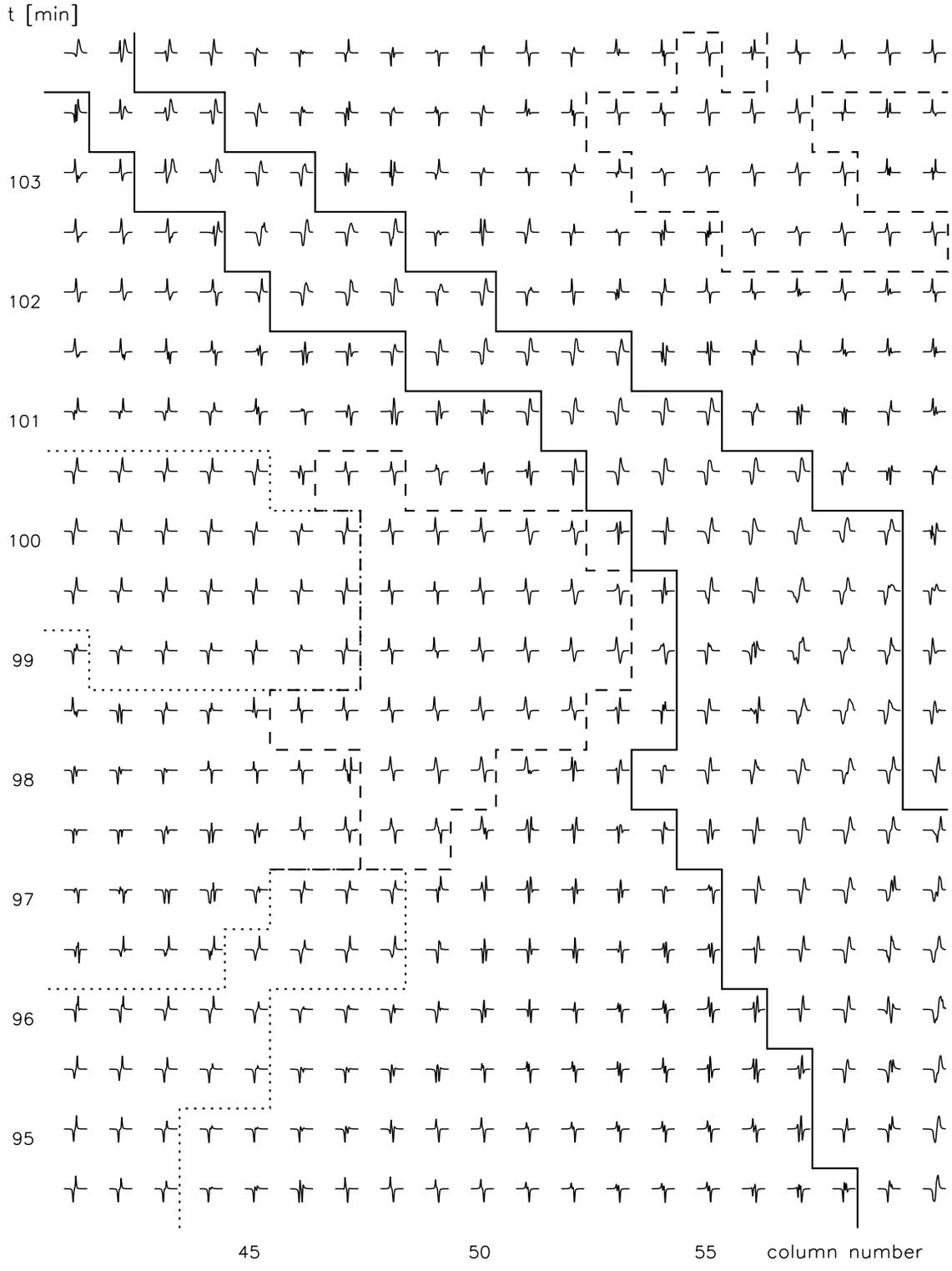}} \hfill
 \caption
{The Stokes $V$ profiles of the Fe I line $\lambda$ 630.2~nm
calculated for every column in a small simulation region fragment
shown by square in Figs 4 and 5; the profiles are normalized to
the amplitude maximum. Solid line) region of a strong flux tube of
negative polarity, dotted line) region of a weak field of negative
polarity, dashed line) positive-polarity field.}
  \label{V-400}
 %}
 \end{figure}
Figure 3 illustrates the $V$ profiles calculated for a small
fragment in the simulation region (square in Figs 4 and 5). Groups
of usual as well as unusual profiles with small Zeeman splitting
can be seen; the profiles with large splitting are marked off by
solid line. They are located in a strong magnetic field region where a
strong flux tube develops. The cases when the profile amplitudes
$a$ and the areas $A$ are equal ($|a_b| = |a_r|$ and $A_b = A_r$)
are very rare in inhomogeneous atmospheres. These equalities are
violated, as a rule, and almost all profiles are asymmetric. One
can see in Fig. 3 that the groups of usual profiles with two
well-defined lobes surround unusual profiles with several lobes or
with a single lobe. At the periphery of flux tubes, profiles with
four or three peak are often found.

We found a wide variety of shapes among the synthesized unusual
profiles. We classified them by the number of lobes rather than by
the profile shape. While the usual profiles have only two lobes, the
unusual ones may have from one to six lobes. The probability $P_n$
that a profile with $n$ lobes will appear depends on limiting
minimal lobe amplitude; we specify this amplitude from the ratio
$r = |a_{min}|/|a_{max}|$ in each profile. Table 1 gives the
occurrence of profiles with different number of lobes at different
limiting minimal lobe amplitudes. Low-amplitude lobes in the
observed profiles are smoothed by atmospheric and instrumental
effects, and we took a sample of unusual profiles with $r = 25$
percent for the further analysis because in this sample the number
of one-lobe profiles is comparable to observation data. All
profiles with two lobes are classed with usual profiles. It is not
improbable that some profiles with lobes of the same sign can be
among them, but their number is insignificant.
%___________________________________ Table 1
%
 \begin{table}[!htb] \centering
 \parbox[b]{14cm}{
 \caption{
Occurrence frequency for profiles with different number $n$ of
lobes and limiting ratio  $r=|a_{min}|/|a_{max}|$. $<a> =
\sum|a_{max}|/N_n$ is the average largest amplitude of each
$n$-group, $P_n$ the fraction of $n$-lobe profiles in the entire
sample (9632), $P_b$ the fraction of blue profiles, $P_{ft}$ the
fraction of profiles formed near flux tubes. \label{T:1} }
\vspace{0.3cm}} \footnotesize
\begin{tabular}{|cccc||ccc||cccc|}
 \hline
     & &  $r=$ 1\%&    &    &$r=$ 15\%&     &   &$r=$ 25\%&&   \\
 \hline
  $n$ & $<a>$ & $P_n$, \% & $P_b$, \% & $<a>$ & $P_n$, \% & $P_{b}$, \%& $<a>$ & $P_n$, \% & $P_{b}$, \% &$P_{ft}$, \%\\
\hline
   1 &0.045&0.01&100 &0.055&0.4&100&0.056  &2   &91& 3\\
   2 &0.104&52&92    &0.094&72  &91&0.091  &79  &90& 10\\
   3 &0.080&21&81    &0.070&18  &74&0.068  &14  &74& 13\\
   4 &0.054&23&83    &0.046&9   &78&0.045  &5   &77& 10\\
   5 &0.044& 3&69    &0.038&0.7 &59&0.047  &0.3 &61& 19\\
   6 &0.040& 0.9&67  &0.130&0.01&66&0.164  &0.03&66& 67\\
    \hline
\end{tabular}
\end{table}
\noindent
%%%%%%%%%%%%%%%%%%%%%%%%%%%%%%%%%%%%%%%%%%%%%%%%%%%%%

According to Table 1, the fraction of unusual profiles in the
sample with $r = 25$~percent is equal to 21~percent, among them
the three-lobe profiles are most often met (14~percent). They are
often similar to the $U$ and $Q$ profiles in shape (Fig.~2a), but
other shapes are also met; for example, all three lobes may have
positive amplitudes. Profiles with four lobes (Fig. 2b) comprise
5~percent, and one-lobe profiles (Fig.~2c) comprise 2~percent.
Other types of profiles (with five and six lobes, Figs~2d--f)
occur more rarely. The lobes in all the profile types illustrated
in Fig.~2 may have other polarities and their amplitudes and
displacements may be different. Profiles with lobes of the same
polarity are quite rare. One-lobe profiles have one well-defined
lobe in one wing, while no lobe can be seen in the other wing or
there can be a relatively small lobe. As can be judged from the
data in Table~1, in most profiles the maximum amplitude is
observed in the blue wing.

The mean greatest amplitude magnitude $<a>$ in the profile groups
for $2\leq n \leq 5$ decreases with growing number of lobes. This
suggests that the unusual profiles are weak, as a rule. The mean
amplitude in one-lobe profiles is somewhat higher than the minimal
amplitude in groups, and the amplitude in the profile groups with
the largest number of lobes is much greater than all other
amplitudes because these groups include, in addition to weak
profiles, very strong ones which are formed near strong flux
tubes. For example, one of two six-lobe profiles in Fig.~2e is
very weak ($a_{max} = 0.004$) while the other in Fig.~2f is very
strong ($a_{max} = 0.4$).

Generally all the unusual profiles which we synthesized can be
divided into two types -- multilobe and one-lobe profiles. In
studies  \cite{27} and  \cite{28} the observed anomalous profiles
were divided into three types -- mixed, dynamic, and one-lobe
profiles. The dynamic profiles differ from the mixed ones by
considerable Doppler shifts only. In our classification they form
a common type of multilobe profiles.

In Table 2 we compare the number of profiles of various types
obtained in the synthesis and from observations. For the analysis
we selected those observed profiles in which the amplitude of the
strongest lobe exceeded 0.0015 in  \cite{27} and 0.001 in
\cite{30}. The calculated profiles have the amplitudes
$|a_{max}|\geq 0.0014$, and they all were used in the analysis.
One can see that our results, on the whole, are not at variance
with the observations. They also are in accord with the profiles
synthesized in  \cite{15} with the use of 3-D MHD models \cite{38}
with a spatial resolution of 20~km. The results of  \cite{15}
suggest that the number of unusual profiles decreases with growing
magnetic flux: as the mean magnetic field strength increases from
0.1 to 14~mT, the number of unusual profiles changed from 35 to 23
percent and the number of one-lobe profiles decreased from 7 to 2
percent.

Note that our earlier results on the one-lobe profiles  presented in
study  \cite{20} have been obtained with the other sequence of 2-D MHD 
models described in \cite{19}. Those MHD models had  a smaller
initial mean strength of bipolar magnetic field (3.2~mT), 
a smaller magnetic flux, and a weaker magnetic flux tubes. Besides,
those models had  a larger simulation region
(5000$\times$1600~km) and a smaller spatial step (25~km). The fraction of
unusual profiles was   larger (45.5 percent of all synthesized profiles) 
than the one in this analysis. The fraction of one-lobe profiles
was also larger (3.6 percent).  
 %%%%%%%%%%%%%%%%%%%%%%%%%%%%%%%%%%%%%%%%%%%%%%%%%%
%___________________________________ Table 2
%
 \begin{table}[!htb] \centering
 \parbox[b]{15cm}{
 \caption
{Number of profiles of various types (observations and synthesis).
\label{T:2} } \vspace{0.3cm}}
 \footnotesize
\begin{tabular}{|l|l|c|c|c|c|}
  \hline
  Data& All observed & Unusual    & $n$-lobe  &One-lobe & One-lobe  \\
      & $V$ profiles &  profiles  & profiles  & ''blue``&  ''red`` \\
  \hline
  Active region \cite{27}&99638 (72\%)&6275 (6.3\%)& 5248 (5.3\%)&660 (0.7\%)&367 (0.3\%) \\
  Quiet region \cite{27}&38360 (37\%)&2599 (6.8\%)& 1698 (4.4\%)&707 (1.8\%) &193 (0.5\%) \\
  Quiet region \cite{30}&9360 (27\%)&3276 (35\%)&--- &--- &--- \\
 Synthesis (this study)  &9632 (100\%) &2050 (21\%) &1863 (19\%) &171 (1.8\%) &16 (0.2\%) \\
   \hline
\end{tabular}
\end{table}

%-------------------------------------------------

\section{Results}
     \label{Results}

{\bf Spatial distribution of unusual profiles over granulation
surface.} Figure 4 displays the spatio-temporal distribution of
unusual profiles on the background of the vertical velocity field
component at the level $\tau_5 = 1$. The data on the velocity
field were taken from the 2-D MHD simulation results. One can see
on the velocity field image the zero velocity line which separates
the upflow regions (light shading) from the downflow regions (dark
shading). This velocity field image can be arbitrarily compared
with the observed Dopplerograms, which also demonstrate the
granulation pattern over the photosphere surface. We do not
present here the traditional granulation pattern as a continuum intensity
distribution because our interest is with the velocity field,
which directly affects the profile asymmetry. The distribution of
the synthesized unusual profiles can be compared to the pattern of
granules and lanes seen in Fig. 4. The unusual profiles are
located in large groups mainly in the border zones of granules and
lanes. They are sometimes met at granule centers. Among unusual
profiles, the one-lobe profiles do not form large groups. They
occur between usual and unusual profiles or among other unusual
profiles are the borders of granules and lanes or between them
near the zero lines of vertical velocities, i.e., at the sites
where the ascending granular flows are replaced by intergranular
downflows.
 %%%%%%%%%%%%%%%%%%%%%%%%%%%%%%%%%%%%%%%%%%% Figure 4
 \begin{figure}
 \centerline{\includegraphics    [scale=1.]{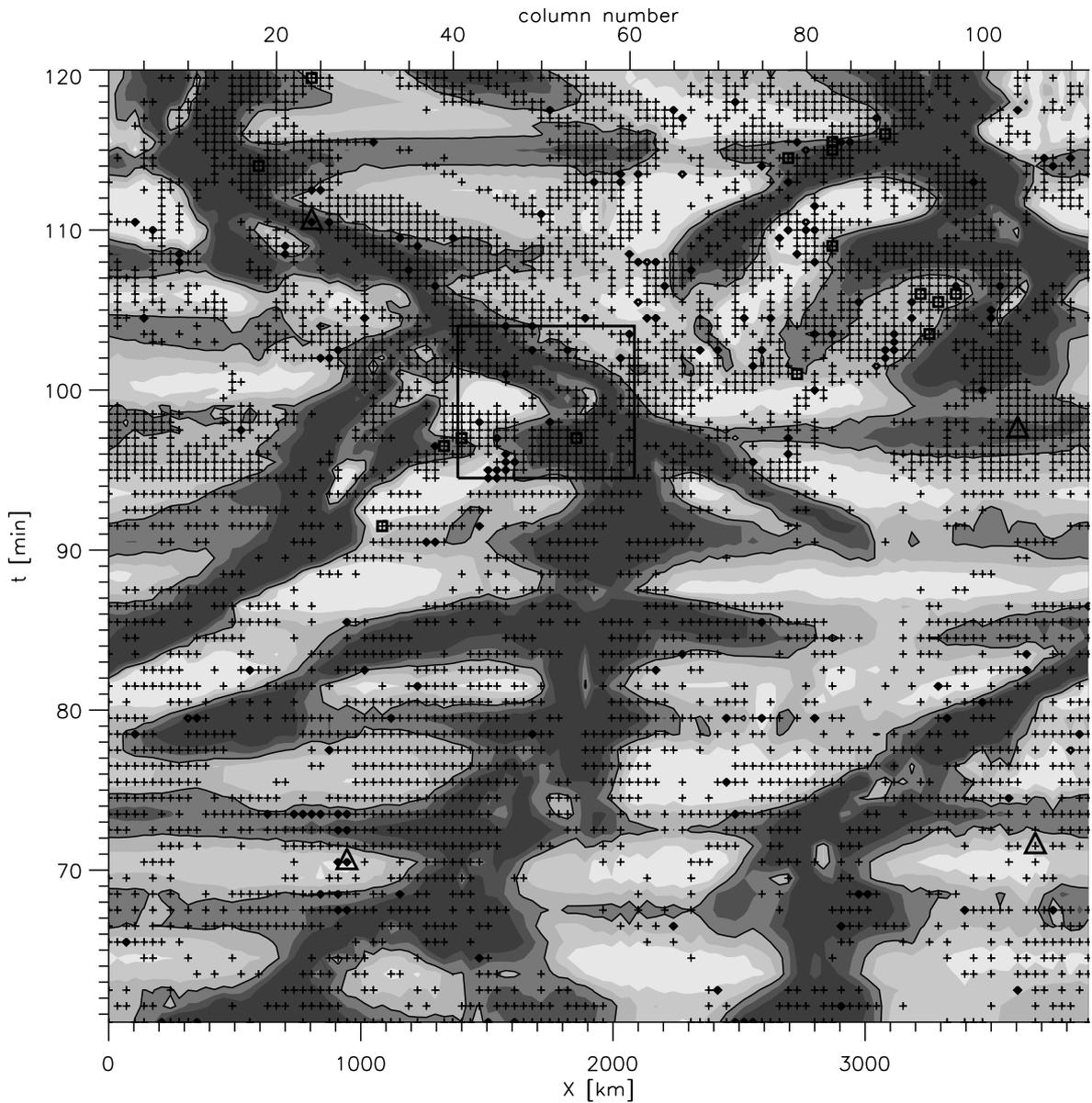}}
 % \hfill
 \caption
{Velocity field distribution in the simulated region at the
$\tau_5= 1$   level, over the course of 60~min. Six shading
gradations from dark (downflows) to light (upflows) correspond to
vertical velocities (in units of km/s) in the intervals 2, 2--1,
1--0, 0--(-1), (-1)--(-2), $<$(-2). Solid line) zero velocity
line. Plus) unusual $V$ profile, square) one-lobe red profiles,
diamond) one-lobe blue profiles, triangle) profiles for which
their formation conditions were analyzed. } \label{VZ}
 %}
 \end{figure}

We also plotted the magnetic field strength distribution (Fig. 5).
The distribution of areas with fields of different polarities is
very nonuniform. Relatively large areas of the same field polarity
are met in granulation regions far from strong flux tubes and
intense downflows. The largest area with the magnetic field of
positive polarity with a strength below 100~mT is outlined by the
coordinates $x_1 = 2200$~km, $x_2 = 3500$~km, $t_1 = 85$~min, $t_2
= 100$~min. There are many mixed-polarity fields, one such area
can be seen beginning with $t=100$~min. At $t = 104$~min and $x =
1300$~km two flux tubes came close together and broke down. These
flux tubes appear in Fig. 5 as sites with the darkest shading.
After the moment $t = 104$~min the field structure became heavily
complicated throughout the simulation region, there appeared a
large number of small areas with opposite polarities and neutral
lines.
 %%%%%%%%%%%%%%%%%%%%%%%%%%%%%%%%%%%%%%%%%%% Figure 5
 \begin{figure}
 \centerline{\includegraphics    [scale=1.]{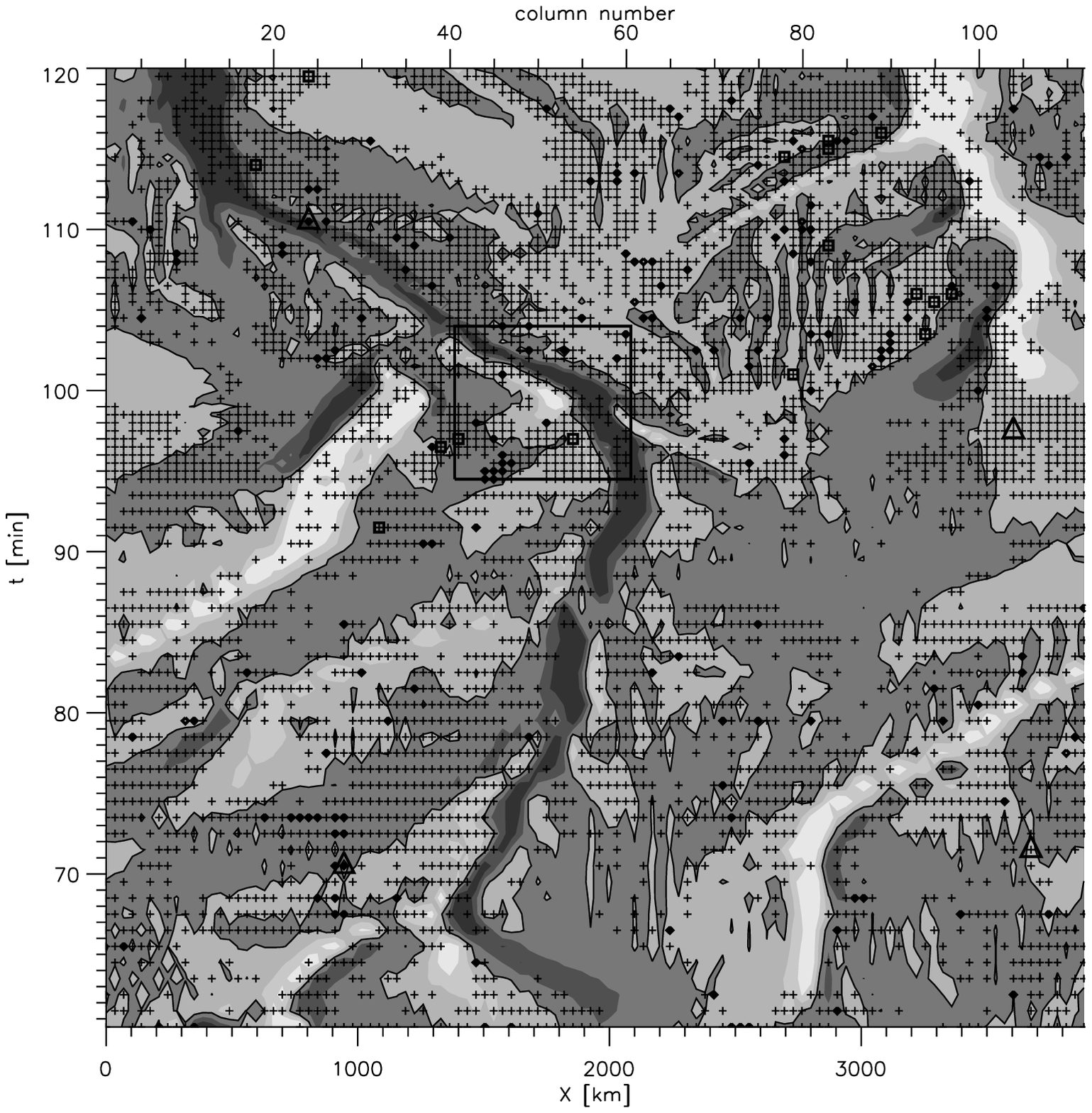}}
 %\hfill
 \caption
{Magnetic field strength distribution at the $ \tau_5= 1$ level.
Light shading) positive polarity, dark shading) negative polarity.
Solid line) neutral magnetic field lines. Field strength
gradations: $>$100, 100--50, 50--0, 0--(-50), (-50)--(-100),
$<$(-100)~mT. Profile shape designations are the same as in
Fig.~\ref{VZ}.
  } \label{BZ}
 %}
 \end{figure}

Now we examine the position of unusual profiles with respect to
the surface structure of magnetic fields. They appear close to
neutral lines as well as far from them. One-lobe profiles are met
almost without exception along neutral lines, where the field
polarity changes. The unusual profiles prefer to cluster in the
areas of weak mixed fields below 70~mT, and 12 percent of unusual
profiles are met close to strong flux tubes ($B_z >70$~mT). It is
evident from Table 1 that the probability for an unusual profile
to appear near a strong flux tube is greater when the number of
lobes in the profile is greater.

We  found that large clusters of unusual profiles are always met
in the areas with intense downflows, where large granules
disintegrate and flux tubes begin to form. An example of such an
area can be seen in Fig. 4 and 5 ($x_1 = 3200$~km, $x_2 =
3900$~km, $t_1 = 95$~min, $t_2 = 99$~min). A strong flux tube
of positive polarity appeared subsequently in this area.

The occurrence of unusual profiles is different in regions with
different magnetic flux -- the number of unusual profiles
decreases, on the average, with growing mean magnetic  field
strength  $<|B_z|>$ in the region (Figs 6a and c). The same tendency was
noted earlier in  \cite{15}. At the same time the number of
unusual profiles tends to decrease when the flux density $<B_z>$
differs from zero (Fig. 6b and c) and when the magnetic field of one or
other polarity begins to dominate in the region. We can say that
this dependence reflects the extent to which fields of different
polarities are mixed in the magnetic region. The closer the
quantity $<B_z>$ to zero, the more pronounced is the alternation
of field polarities in the region (cf. Figs 5 and 6).
  %%%%%%%%%%%%%%%%%%%%%%%%%%%%%%%%%%%%%%%%%%% Figure6
 \begin{figure}
 \centerline{\includegraphics [scale=1.]{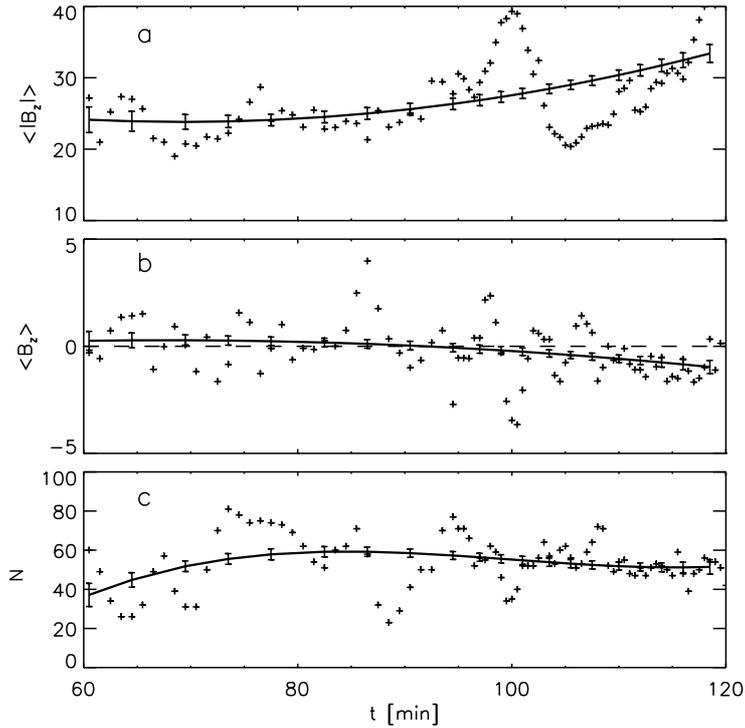}}
% \hfill
 \caption
{Temporal changes in the simulation region: a) mean field strength
$<| B_z |>$ at the $ \tau_5=1$ level, b) magnetic flux density $<
B_z>$ at the same level, c) number $N$ of unusual profiles.
  } \label{BZ-N}
 %}
 \end{figure}

So, when we compare the distribution of unusual profiles over the
surface with the distributions of velocity, magnetic field
strength, and field polarity, it becomes apparent that the
presence of mixed-polarity fields and neutral lines is a very
important prerequisite to the appearance of unusual profiles. The
one-lobe profiles, unlike the multilobe ones, prefer the sites
near the neutral line and are often met near the zero line. The
fact that the one-lobe profiles are often met at the very edges of
granules and lanes rather than in their central parts indicates
that they arise under most extreme conditions in passing from one
structures to others. The asymmetry of one-lobe and multilobe
profiles seems to be of the same origin. To elucidate this
problem, we consider in greater detail, with invoking the
contribution functions, the formation of these profiles and the
effect of the magnetic field and velocity field gradients on their
shapes.

%%%%%%%%%%%%%%%%%%%%%%%%%%%%%%%%%%%%%%%%%%%%%%%%%%%%%%%%%%%
%___________________________________ Table 3
%
 \begin{table}[!htb] \centering
 \parbox[b]{12cm}{
 \caption
{Distribution of one-lobe profiles over granulation surface.
\label{T:3} } \vspace{0.3cm}}
 \footnotesize
\begin{tabular}{|l|c|c|c|c|}
  \hline
  Type & General number& Granular & Intergranular & Edge \\
  \hline
 One-lobe profiles &187 &82 (43\%)&61 (33\%)&44 (24\%) \\
 One-lobe ''blue`` profiles    &109 &69 (63\%)&23 (21\%)&17 (16\%) \\
 One-lobe ''red``profiles  & 78 &13 (17\%)&38 (49\%)&27 (34\%) \\
    \hline
\end{tabular}
\end{table}
{\bf One-lobe profiles.} As would be expected from observation
data  \cite{27}, we found two types of synthetic one-lobe
profiles: 58 percent of ''blue`` profiles and 42 percent of
''red`` ones (see an example of ''blue`` profiles in Fig. 2c). We
also distinguish granular, intergranular, and marginal one-lobe
profiles, in accordance with Fig. 4. Table 3 gives the statistics
of the distribution of unusual profiles. The occurrence of
one-lobe profiles is greater in the granular regions, and they are
predominantly ''blue`` there (84 percent), while in the lanes and
at the granule-lane borders ''red`` profiles are met most often
(60 percent). Such statistics suggests that the granulation
velocity field and velocity gradients have a considerable impact
on the formation of one-lobe profiles.

 %%%%%%%%%%%%%%%%%%%%%%%%%%%%%%%%%%%%%%%%%%% Figure 7
 \begin{figure}
 \centerline{\includegraphics    [scale=0.95] {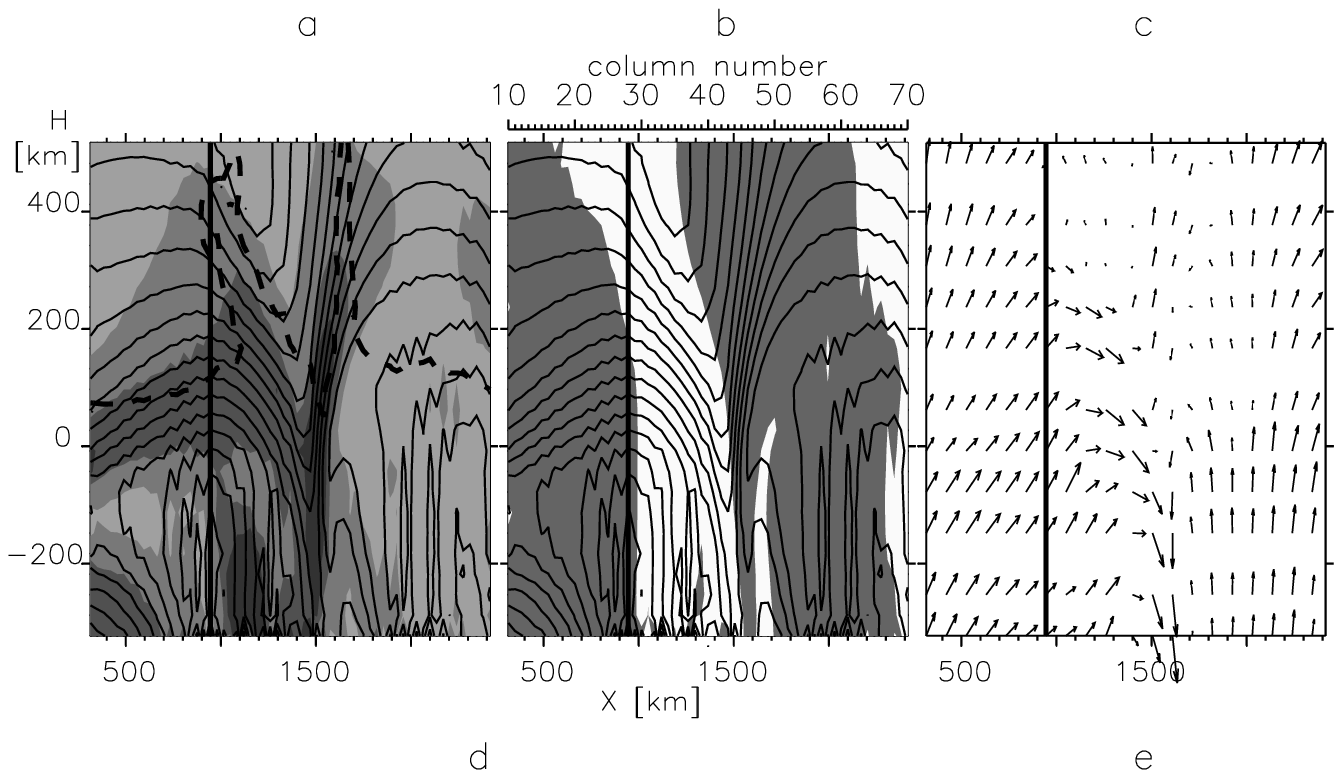}}
 \centerline{\includegraphics    [scale=0.8]{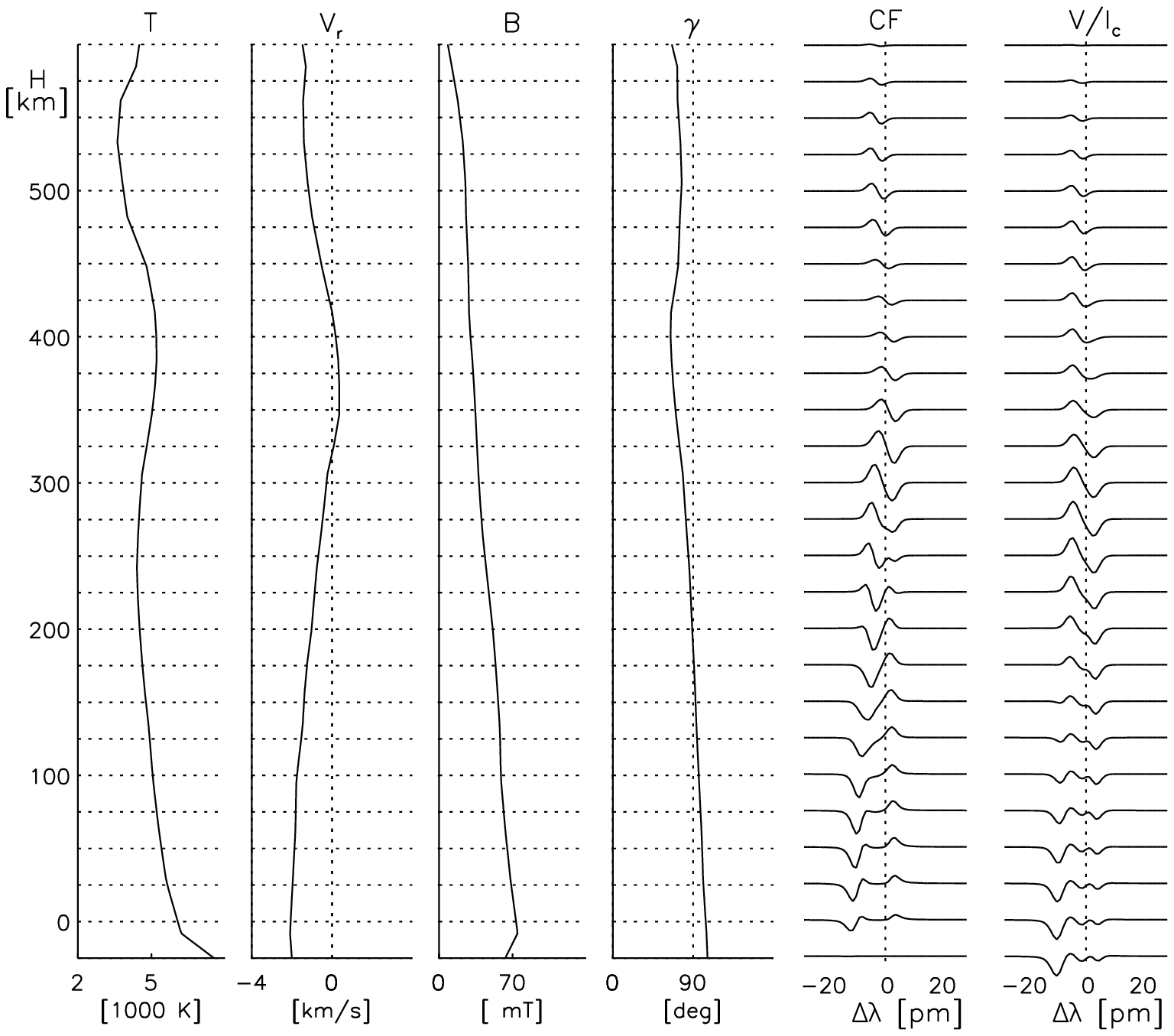}}
 % \hfill
 \caption
{Formation of a granular $V$ profile with blue wing: a)  snapshots
of magnetic field strength for the simulation region
fragment (four shading gradations from dark to light correspond to
70, 70--50, 50--25, 25--0~mT); b) polarity: lighter shading)
positive, darker shading) negative; c) velocity field. Thin line)
magnetic field lines, thick line) the line of sight along which
the profile is calculated, Thick dashed line) level where $\lambda
= 5000$~K; d) vertical profiles of temperature $T$, velocity
$V_z$, magnetic field strength $B$ and inclination $ \gamma$; e)
profiles of depression contribution functions ($CF$) and emerging
relative profile intensity ($V/I_c$) at various levels in the
atmosphere. } \label{V-lobe}
 %}
 \end{figure}
Now we consider the formation of ''blue`` profiles with the use of
calculated depression contribution  functions  along the line of
sight. Figure 7 illustrates the formation of one of the granular
''blue`` profiles (triangle in Figs 4 and 5, $t= 70.5$~min, $x =
980$~km, column 28). The upper panel shows the vertical
snapshot of a simulation region fragment; the thick solid vertical
line is the line of sight along which the selected profile is formed. One
can see from the distributions of the field strength (Fig. 7a),
field polarity (Fig. 7b), and velocity field (Fig. 7c) that this
specific line runs at a distance of 500~km from the center of a
magnetic flux tube, it crosses a nearly horizontal magnetic field
of negative polarity in deeper layers and positive polarity in
upper layers. The field inclination increases in the upper layers
due to the magnetic canopy effect. The vertical velocity of
ascending flow decreases with increasing height from 2 km/s to
zero at the canopy boundary. The vertical profiles of temperature
T, velocity $V_r$, field strength $B$, and field direction
$\gamma$ corresponding to the line of sight are plotted in Fig.
7d. Figure 7e displays the depression contribution function
profiles $CF$($ \lambda$) at the specified geometric heights $H$
as well as the $V$ profiles obtained by integrating $CF$ along the
heights above the given level, i.e., $V/I_c(\tau,\lambda) =
\int_\tau^0  d\tau_0 CF(\tau_0 )$. One can see \cite{5} for more
information on the calculation of contribution functions $CF(
\lambda )$. At the very bottom of the right panel in Fig 7e one
can see the one-lobe absorption $V$ profile of the circularly
polarized light emerging at the surface.

When the profiles are compared, it is apparent that the level of
effective formation of the line $\lambda$~630.2~nm profile lies in
the temperature region about 5000~K. The position of this
effective region in the inhomogeneous atmosphere is shown in Fig.
7a by a dashed thick curve on the background of the magnetic field
strength distribution. The depression contribution in the profile
of emerging polarized radiation are the greatest in every column
near this temperature layer. It is significant that temperature
rises near flux tubes. This rise is manifested the so-called hot
walls which always appear in the upper flux tube layers due to
magnetoconvection  \cite{3,9}.

Considering the vertical velocity  profile ($V_r$)  in a given column (Fig. 7d)
and the velocity pattern in the surrounding space (Fig. 7c), one
can  notice a flow which is directed towards the flux tube
and which fosters the field concentration in the flux tube. The
profile we consider here is formed in the region where the
velocity gradient changes its sign from negative to positive when
going from deeper layers to upper ones. As a result, the blue
Doppler shifts of the contribution function profiles are different
at different photospheric levels so that the integrated profile is
strongly shifted to shorter wavelengths. Thus, a sudden drop of
velocity with growing height gave rise to a stronger blue wing in
the profile.

As the magnetic field strength decreases with increasing height
(Fig. 7d), the magnetic splitting becomes weaker and the distance
between the contribution profile lobes decreases, while the change
in the magnetic field direction along the line of sight reverses
the shape of the contribution function. Due to such intricate
combinations of contribution functions, the intensities in the red
profile wing compensate one another. The cause of the polarity
change along the line of sight is evident from Fig. 7b -- this is
a kink of magnetic field lines in the region of the ascending
convective flow which carries the horizontal magnetic field.

Thus, when all the above-mentioned effects are integrated over all
layers, the contributions with opposite signs are compensated in
the red wing and are accumulated in the blue wing. The major
contribution to profile formation is made in the lower
photospheric layers, where the magnetic field has the positive
polarity. As a result, the integrated profile has a blue wing with
negative amplitude. The polarity change and the negative field
strength gradient in the region of profile formation suppress the
red wing, while the negative gradient of vertical velocity
enhances the blue wing. Our analysis of the formation of a large
number of blue one-wing profiles revealed that the negative
velocity gradient along the line of sight and the polarity change
are typical for them, while the field strength gradient can be
negative as well as positive. The field strength gradient along
the line of sight can also change its sign. Therefore, the
presence of blue profiles in granules suggests that the field
polarity changes its sign there, there are significant negative
gradients of vertical velocities of upflows, and that various
gradients of the weak magnetic field strength can be met. The sign
of wing amplitude is indicative of the field polarity in the layer
most efficient in the formation of the given profile.

The conditions for the granular red profile formation differ from
those for blue profiles in that the upflow velocity gradient is
predominantly positive, i.e., the velocity amplitude increases
with height (Fig. 8d).
 %%%%%%%%%%%%%%%%%%%%%%%%%%%%%%%%%%%%%%%%%%% Figure 8
  \begin{figure}
 \centerline{\includegraphics    [scale=1.]{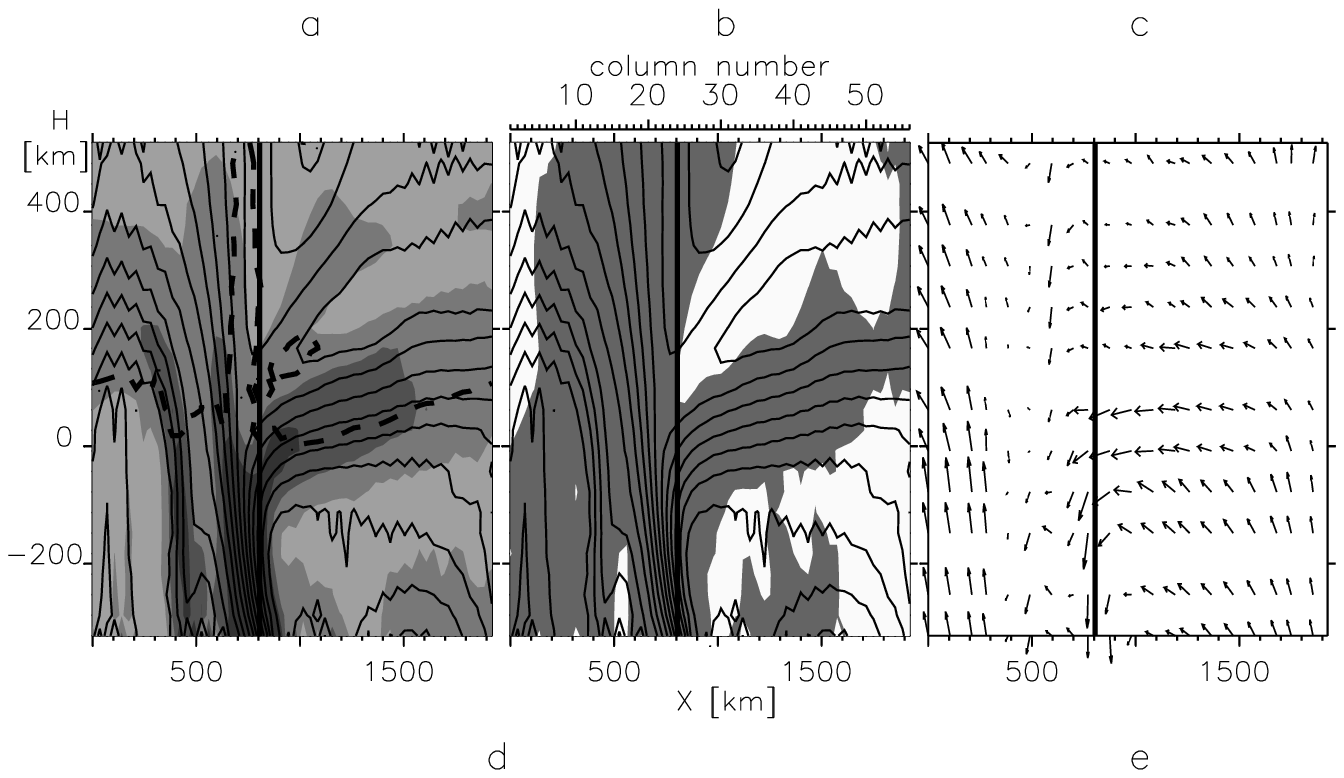}}
 \centerline{\includegraphics    [scale=0.85]{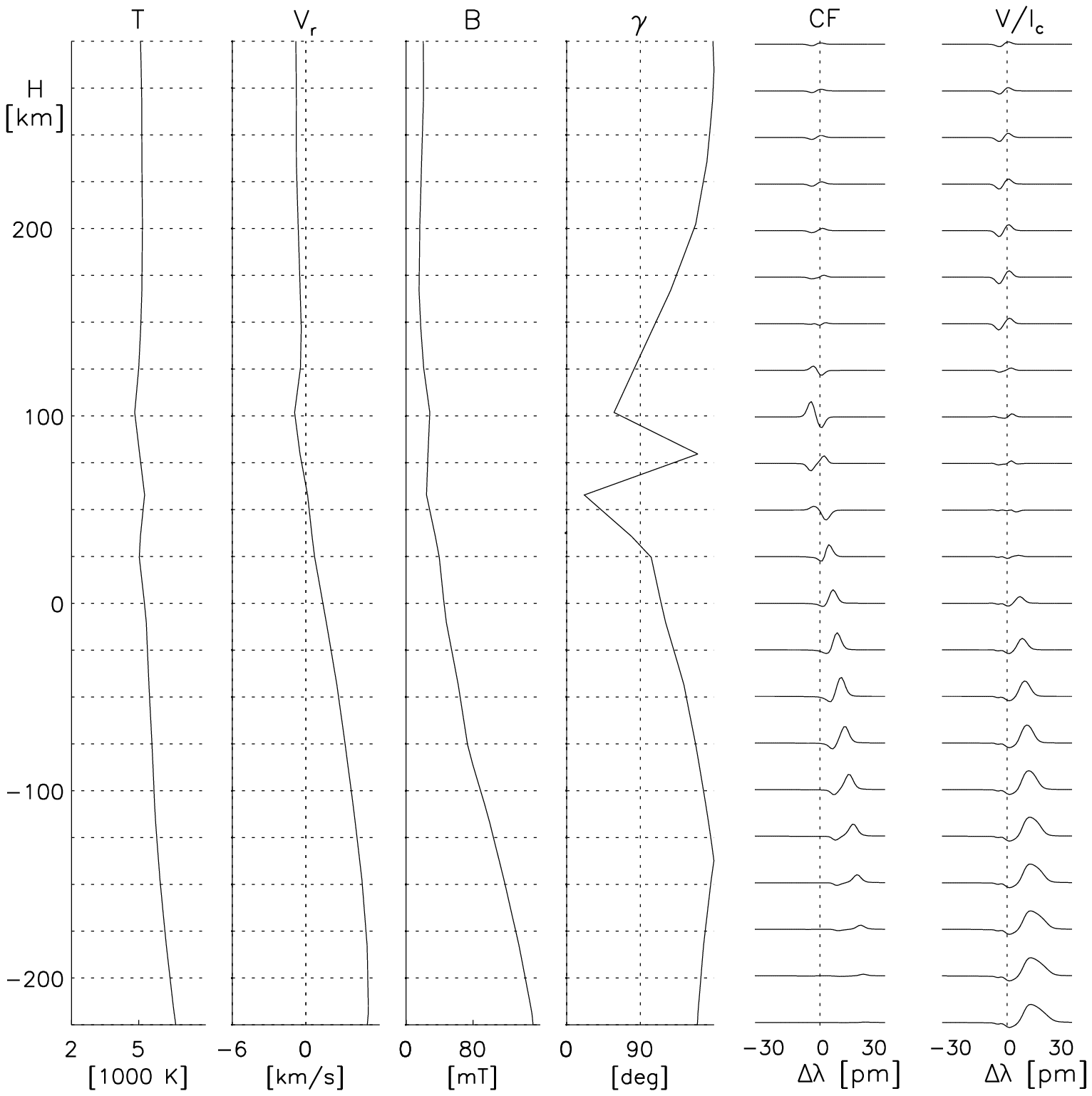}}
 \caption
{Formation of an integranular $V$ profile with red wing near a
strong flux tube (designations are the same as in Fig. 7).
  } \label{V3}
 %}
 \end{figure}
The red profiles are formed in the intergranular lanes with weak
magnetic field when downflow velocities in the effective formation
region exceed 2 km/s. When the velocities are lower, the
probability for a blue wing only to be formed is much higher. If
there are strong flux tubes in the lane, the red profiles are
formed more often than the blue ones on their periphery. 
Figure 8 illustrates the formation of an intergranular red profile on the
periphery of a strong flux tube ($x = 840$~km, $t = 110.5$~min,
column 24 in Figs 4,5). Steep negative gradients of downflow
velocity and field strength can be seen in the effective profile
formation region. There also is a small region when the polarity
changes its sign. As a result, a quite wide red wing is formed and
the blue wing is suppressed.

Red peripheral one-wing profiles as well as blue ones located on
the zero line and on the neutral line between granules and lanes
are formed in the regions where the velocities are close to zero
and their gradients are insignificant so that the gradients of
magnetic field strength and direction have a dominant role there.

Our analysis revealed that the one-wing profiles are formed
only when the magnetic field polarity changes along the
line of sight and when there are the velocity and field strength
gradients. So, the appearance of one-wing profiles points
unambiguously to the presence of inhomogeneous magnetic fields of
different polarities in the photosphere. The shape of these
profiles is difficult to use for the diagnostics of physical
conditions in magnetic structures, because there are various
factors which can affect this shape. First of all, this is a
horizontal nonuniformity of the solar surface structure \cite{15}.
In this analysis we investigated the profile asymmetry without 
averaging and found
that structure irregularities can occur in inhomogeneous models
along the line of sight as well, and they heavily affect the shape
of the Stokes profiles. We also found that the shape of one-wing
profiles depends on their location on the granular surface of the
photosphere. We may state, therefore, that the predominance of
blue one-wing profiles in the region observed attests to the
domination of almost horizontal weak granular magnetic fields. The
predominance of red profiles can testify that there are mixed
strong and weak magnetic fields and considerable field strength
and velocity gradients in the region.

{\bf Multilobe profile.} We used the contribution functions to
analyze a large number of multilobe profiles and the conditions of
their formation. It has been ascertained that in all instances the
polarity changed one or more times along the line of sight
whatever the profile located may be -- near the neutral line, far
from it in a granule, or in a lane at a flux tube periphery.
Gradients of velocity and field strength also are requisite, with
the gradient sign changing very often. The greater the number of
lobes in the profile, the more intricate are the gradient
profiles.

%%%%%%%%%%%%%%%%%%%%%%%%%%%%%%%%%%%%%%%%%%% Figure 9
  \begin{figure}
 \centerline{\includegraphics    [scale=1.0]{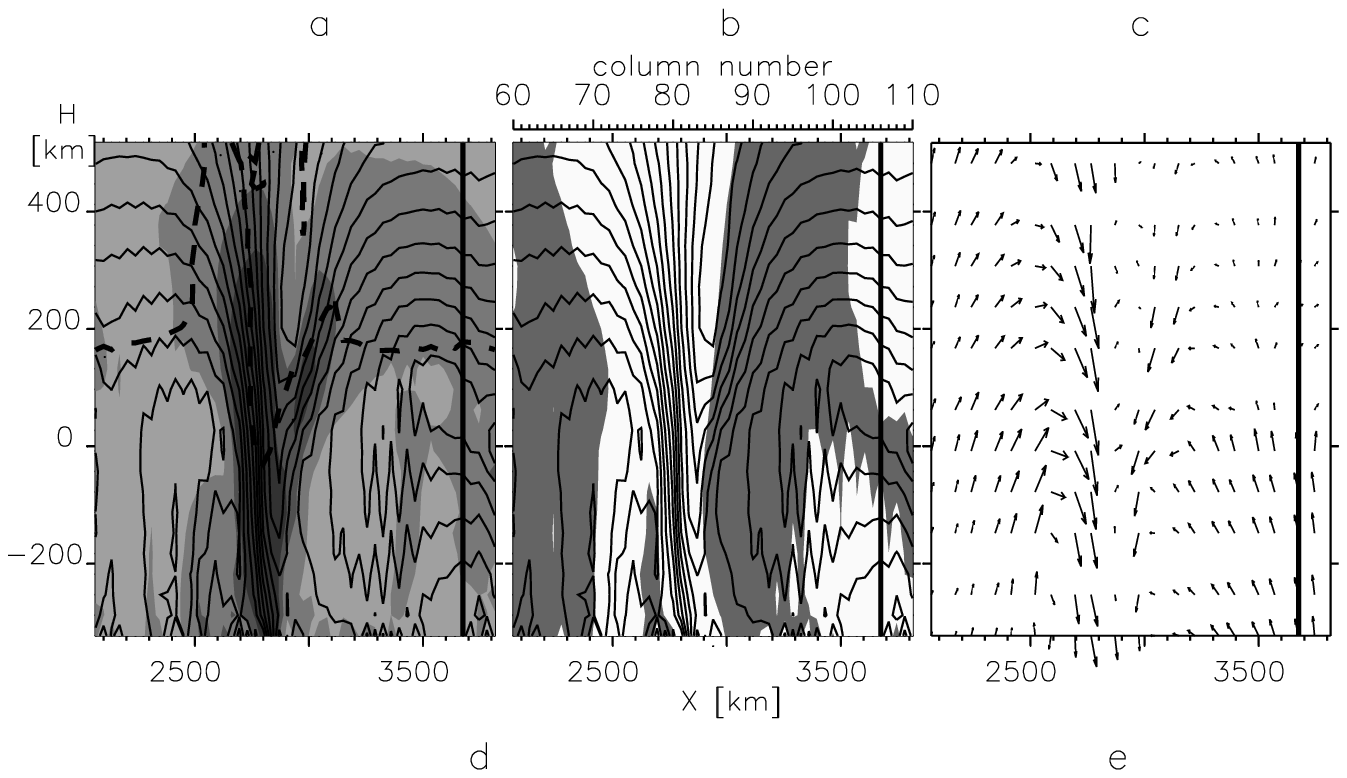}}
 \centerline{\includegraphics    [scale=0.85]{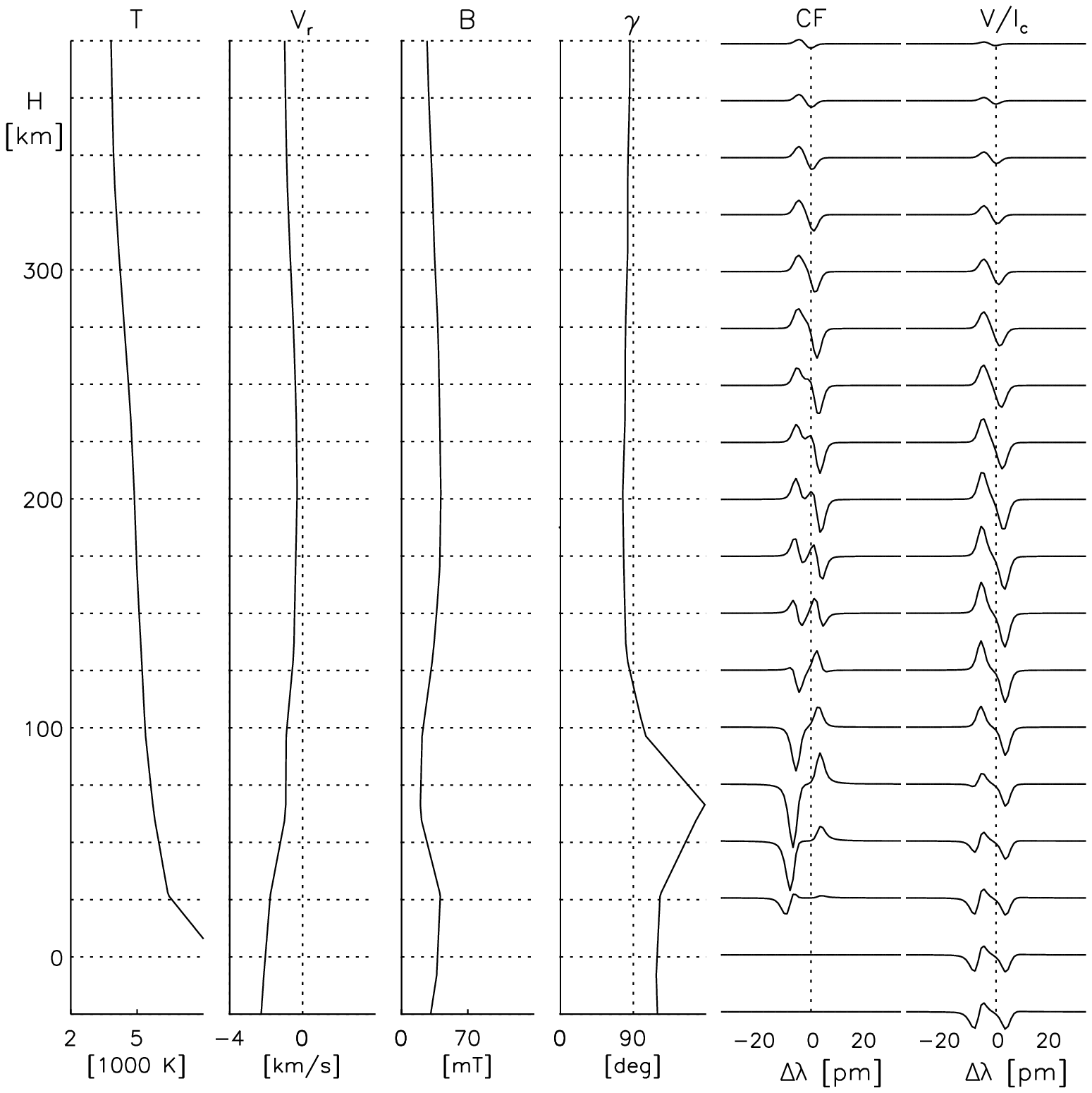}}
 \caption
{Formation of a typical unusual $V$ profile with three lobes
(designations are the same as in Fig.7).
 }
 \end{figure}
 %%%%%%%%%%%%%%%%%%%%%%%%%%%%%%%%%%%%%%%%%% Figure 10
 \begin{figure}
 \centerline{\includegraphics    [scale=1.]{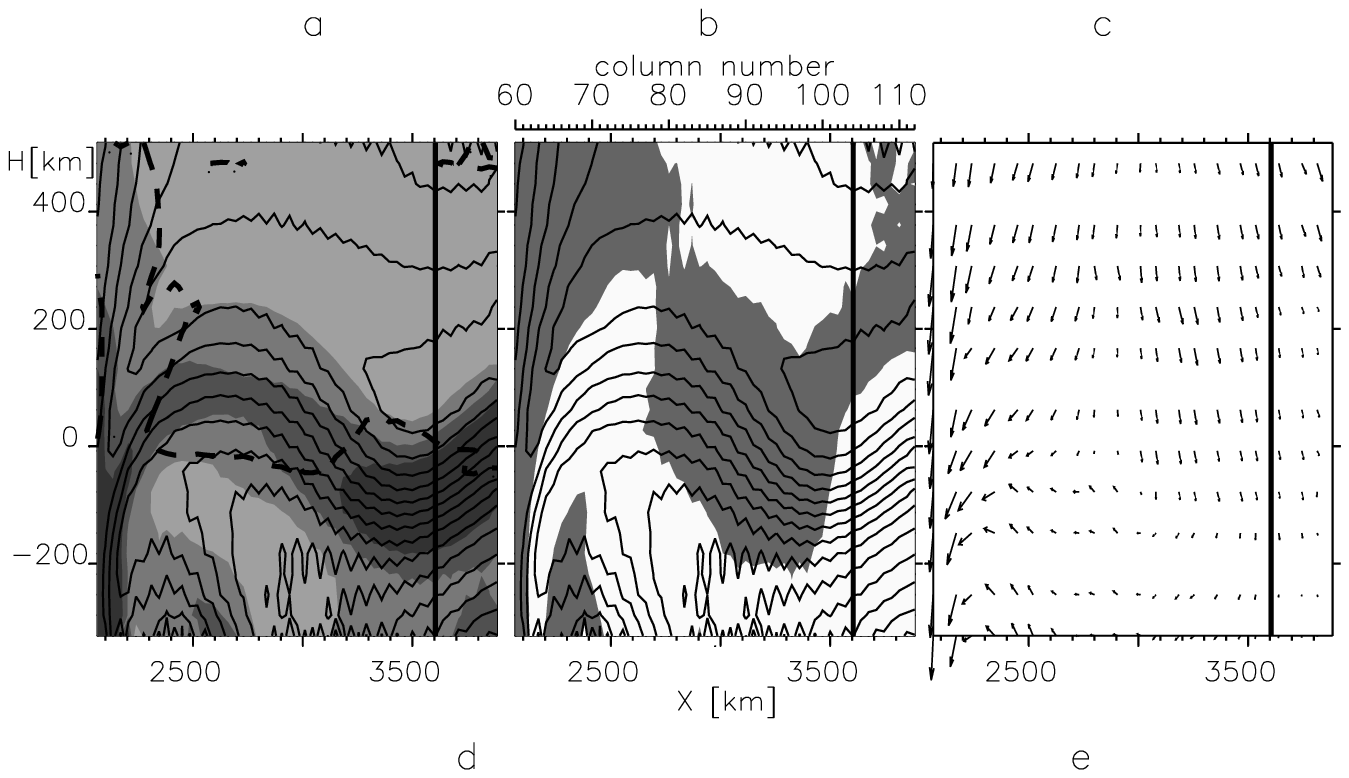}}
 \centerline{\includegraphics    [scale=0.85]{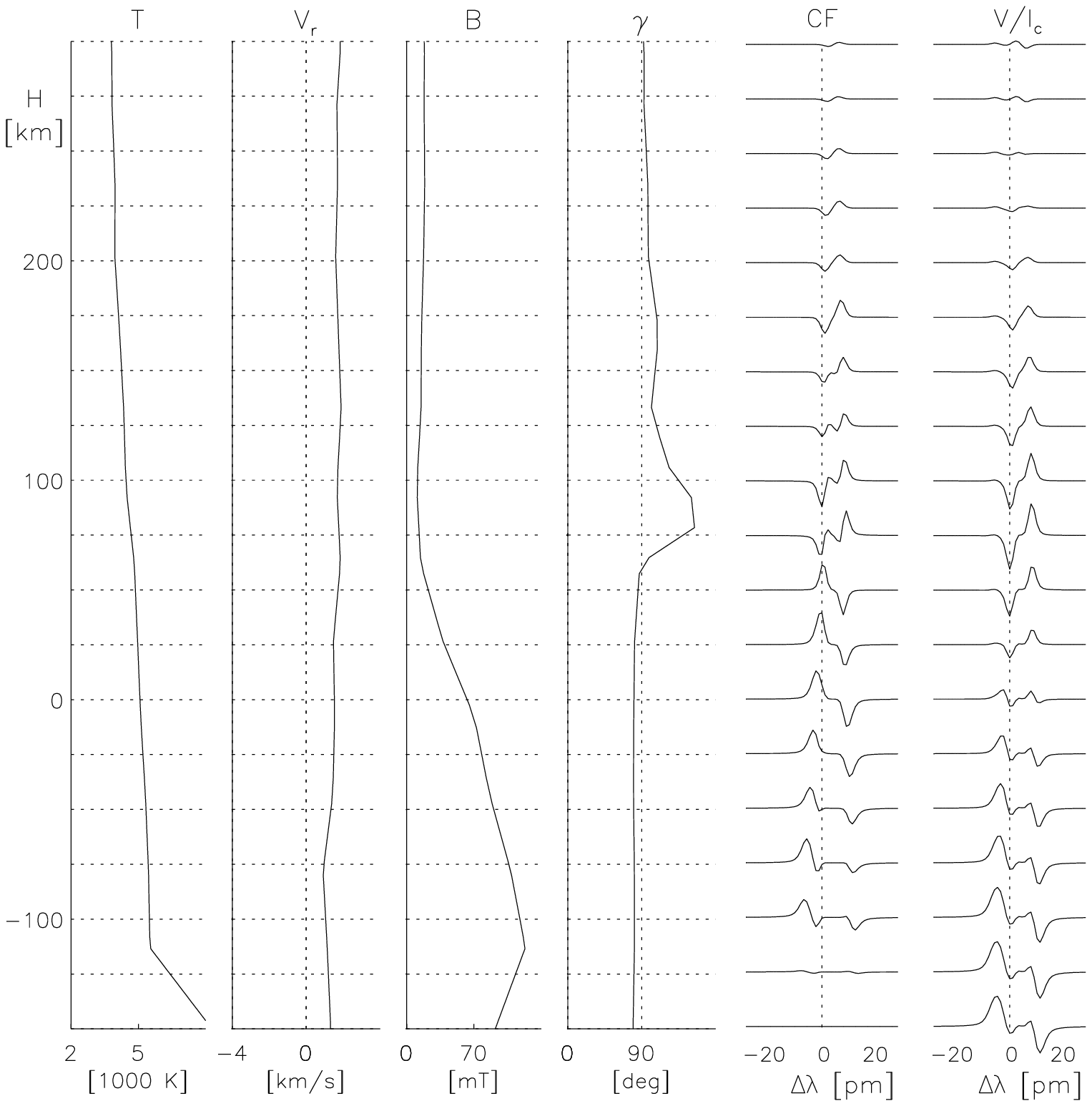}}
 \caption
{Formation of a typical unusual  $V$ profile with four lobes
(designations are the same as in Fig. 7).
  } \label{V10}
 %}
 \end{figure}
Figure 9 illustrates the formation of a typical three-lobe
profiles which is often met in observations. Column 106 (see Figs
4,5), along which the profile was calculated, is located far from
a flux tube in an upflow ($x = 3710$~km, $t = 71.5$~min). We
purposely chose this example so that the velocity gradient might
be similar to the gradient considered earlier in the blue one-wing
profile in Fig. 7. Nevertheless, the magnetic field gradient is
different in these two profiles. Here the field strength decreases
with growing height in the profile formation region, and then it
increases. It is precisely this gradient sign change that produces
a three-lobe profile.

Four-lobe profiles are also often observed. An example of such a
profile formed in a stable downflow is seen in Fig. 10 ($x =
3640$~km, $t= 97.5$~min, column 104 in Figs 4,5). Fragmentation of
a large granule and formation of a flux tube begin in this region
(Figs 10a--c). These processes are described in detail in
\cite{4,9}. The vertical profiles of atmospheric parameters (Figs
10d, e) indicate that there are two factors which affect the
profile shape and give rise to four lobes: ($i$) the strong
horizontal magnetic field of about 100~mT which is concentrated in
deep photospheric layers and which becomes weaker with increasing
height and then remains almost unchanged (this situation is
similar to a magnetopause) and ($ii$) the polarity change in the
magnetic field carried by an intense plasma flow into deeper
layers. The vertical velocity of the downflow has a small gradient
and slightly affects the profile shape.

The number of lobes depends on the combination of all asymmetry
factors in each specific case. The formation of multilobe profiles
is related to more intricate structures of velocity field and
magnetic field and to more frequent gradient sign changes. For
these profiles it is difficult to find necessary combinations of
all three parameters (velocity, field strength, and field
inclination) and their gradients. Besides, we do not see any
considerable difference between the mechanism of formation of
one-lobe and multilobe profiles. They all arise due to polarity
changes and changes in the sign of the velocity and field strength
gradients. 

Thus, the main cause of extreme asymmetry is the growth of
magnetic field structurization, which occurs most often not far
from intense matter downflows, where flux tubes are being formed,
or near already existing strong flux tubes.

%-------------------------------------------------

\section{Discussion and conclusion}
     \label{Discussion}

We synthesized 9632 Stokes $V$ profiles based on the inhomogeneous
model atmospheres obtained in the numerical 2-D solar
magnetogranulation simulation  \cite{3}. The highest amplitude of
the wing in the weakest calculated profile was $|a_{max}| =
0.0014$. The number of extremely asymmetric profiles with the
number of lobes smaller and greater than two depends on the limit
chosen for the amplitude of the minimal lobe. This limit can be
taken as one percent for the calculated profiles. To compare our
results to observations, we took a limiting amplitude ratio of 25
percent. With this value, the fraction on calculated one-lobe
profiles corresponded to observations  \cite{27}. The unusual
profiles made up 48 and 21 percent of all calculated profiles when
the limiting amplitude ratio was 1 and 25 percent, respectively.
The results of the  observation  \cite{30} and of the synthesis
with 3-D MHD models  \cite{15,23} suggest that unusual profiles
comprise about 35 percent in the profile sample with $|a_{max}|$
greater than 0.0015 or 0.001. The comparison of our results with
observations is not sufficiently correct because the number of
unusual profiles found in observations was not referred to all
observed profiles. There may be some discrepancy caused by
differences in the magnetic fluxes and by the use of a
two-dimensional approximation in the MHD models. The granulation
motions of matter in the 2-D simulated space necessarily differ
from actual motions, and this reflects on the magnetic field
structure and profile shapes.

The synthesized extremely asymmetric $V$ profiles demonstrated
some peculiarities found earlier in the polarimetric observations
outside active regions made with a spatial resolution better
than one second of arc  \cite{14,27,30}. Our simulation also
revealed that unusual profiles are often grouped near neutral
lines and at the periphery of regions with flux tubes, avoiding
their central parts. The shapes and shifts of these unusual
profiles can be different, the profiles with three and four lobes
being of most frequent occurrence.

Our analysis with the use of depression contribution functions
confirmed the conclusions made in  \cite{15,20,23} that the
one-lobe Stokes $V$ profiles are formed in the solar photosphere 
due to the joint action of the gradients of velocity fields
and magnetic fields of mixed polarities, with the polarities
changing not only on the surface bat along the line of sight as
well. The effect of these polarity changes cannot be detected in
direct observations, it became evident only in simulations, and it
points to a complicated structure of magnetic fields caused by
granulation motions.

The new principal results of our investigation of synthesized
multilobe profiles are as fallows.

1. Unusual profiles appear predominantly near intense plasma
downflows (with velocities of 2~km/s and more) at the borders of
granules and lanes.

2. The large groups of unusual profiles observed in the regions
with weak fields and intense downflows can be precursors of strong
magnetic concentrations in the form of flux in these regions.

3. The one-lobe profiles with well-developed blue wing appear in
granules in the main, while in lanes and near the zero line only
the profiles with red wing appear.

4. All extremely asymmetric profiles are of the same nature
whatever the number of lobes and their shape. The major
prerequisite for the formation of unusual profiles is the change
of magnetic field polarity along the line of sight (there may be
one change or several ones). The gradients of matter velocity, of
magnetic field strength and inclination, which are the main cause
of usual classical asymmetry, still remain important factors in
the formation of unusual profiles, and the number of profile lobes
is proportional to the frequency of changes in the sign of the
magnetic field strength gradient along the line of sight. Various
combinations of all these factors as well as the variability of
main atmospheric parameters in the region of profile formation
give rise to a great diversity of profile shapes.

Our results give answers to some questions raised in  \cite{27} as
to the origin of unusual profiles.

{\it Are there most frequently met types of unusual profiles?} We
found two such types of unusual profiles -- with three and four
lobes of alternating signs.

{\it What physical processes are responsible for the most
frequently met types of unusual profiles?} The turbulent and
convective motions of plasma and the formation of magnetic
concentrations during the fragmentation of big granules are
instrumental in producing a complicated structure of magnetic
fields with sign-variable gradients of field strength and
inclination.

{\it Are one-lobe profiles the extreme case of unusual profiles?}
They might be reckons as such from the standpoint of profile
shape, but as far as the origin of unusual profiles is concerned,
the profiles with four and more lobe should be regarded as the
extreme case of asymmetry.

{\it Are the unusual profiles a result of mixed polarities on very
small scales?} It is precisely the polarity changes along the line
of sight that are the main cause of extreme profile asymmetry.

{\it Why these unusual profiles have never been observed inside
strong kilogauss fields?} Strong magnetic flux tubes with
field strength of more than 100~mT are almost without exception
vertical in the region of profile formation, and no polarity
changes occur in this case.

{\it Can profile shapes be used for preliminary diagnostics of
observed solar surface areas?} The appearance of unusual profiles
is indicative of a complicated magnetic field structure in both
the horizontal direction and the vertical one, but they say
nothing about the field strength. The magnetic fields are often
weak and nearly horizontal in the regions of ascending magnetic
flux. Sometimes they can be inclined in the magnetic canopy
regions not far from flux tubes, and they can be nearly vertical
near kilogauss flux tubes.

{\it What is the cause of mixed polarities? What physical
processes lead to polarity mixing on very small scales?} This is
the principal problem, and it needs a more detailed discussion.

There is no doubt that the extreme asymmetry of $V$ profiles with
one or several lobes is dictated by the fine magnetic field
structure produced by turbulent plasma motions in the regions
where the convection penetrates into the photosphere (the
magneto-convection processes). Three-dimensional hydromagnetic
simulations  \cite{36} demonstrated the existence of a complicated
surface granulation structure and the crucial importance of
intense downdraft in the solar convection. It is these downdraft
that control much more extensive ascending convective flow and
smaller turbulent motions. The authors of  \cite{36} sketched the
following broad outlines of solar convection. The ascending 
warm convective flows have low density and medium
entropy. While ascending, they spread in order to conserve mass,
since the density is lower in the upper layers. Only a small part
of ascending material reaches the surface, where its energy and
entropy are carried away by radiation; thus the material density
increases, and this heavier material forms a downflow. This
descending material forms intensive downdrafts in the lower
layers, obeying the law of mass conservation; shifting motions
arise at the downflow borders, and they produce eddies and
turbulence. To the contrary, in the ascending matter the level of
fluctuations is very low because the spreading smooths all
disturbances. As a result, we observed laminar upflows and
turbulent downflows. When we supplement this picture with a
magnetic field, which is frozen in the plasma under the solar
plasma conditions, we can easily understand why magnetic field
lines have complicated configurations at the borders of granules
and lanes.

The relationship between magnetic field and convection was
demonstrated in two-dimensional MHD simulations of
magneto-convection  \cite{1,3,4,9,35} as well as in
three-dimensional ones  \cite{37,38}. Observations  \cite{14,29}
and the results of numerical simulations of the Stokes profiles
 \cite{7,15,6} also pointed to a strong interdependence of magnetic
fields in quiet regions and local convection processes.

In this study, which is based on the results of the Stokes profile
synthesis and the 2D-MHD simulations of nonstationary
magneto-convection, we demonstrated that the main cause of the
formation of unusually shaped profiles is the complicated
structure of the magnetic field with its polarity changing along
the line of sight and the complicated gradients of velocity and
field strength at granule and lane borders. Such conditions
favorable for the formation of unusual profiles arise due to the
local turbulence produced in the surface magneto-convection by
shifting motions at the borders of powerful downflows.

{\bf Acknowledgements.} The author wishes to thank S. Solanki, S.
Ploner, M. Sch\"{u}ssler, and E. Khomenko for their useful
comments and discussions of the results. This study is a part of
an international program, it was partially financed by INTAS
(Grant No.00084).

%========-------------------------------------------------

%%%%%%%%%%%%%%%%%%%%%%%%%%%%%%%%%%%%%%%%%%%%%%%%%%%%%%%%%%%%

\end{document}